\documentclass[%
 aip,
 hyphens,
 amsmath,amssymb,
 reprint,%
]{revtex4-1}

\usepackage{graphicx}
\usepackage{dcolumn}
\usepackage{bm}

\usepackage[utf8]{inputenc}
\usepackage[T1]{fontenc}
\usepackage{mathptmx}
\usepackage{etoolbox}
\usepackage[]{amsmath}
\usepackage{mathrsfs}
\usepackage{mathtools} 
\usepackage{enumerate}
\usepackage[usenames,dvipsnames,svgnames,table]{xcolor}
\usepackage{IEEEtrantools}
\usepackage{graphicx}
\usepackage{array}
\usepackage{multirow}
\usepackage{verbatim}
\usepackage{soul}
\usepackage{tabularx}
\usepackage{booktabs,siunitx}
\usepackage[bookmarks=false]{hyperref}

\usepackage{physics}
\usepackage{siunitx}
\usepackage{makecell}
\usepackage{float}

\makeatletter
\def\@email#1#2{%
 \endgroup
 \patchcmd{\titleblock@produce}
  {\frontmatter@RRAPformat}
  {\frontmatter@RRAPformat{\produce@RRAP{*#1\href{mailto:#2}{#2}}}\frontmatter@RRAPformat}
  {}{}
}%
\makeatother
\begin{document}

\preprint{AIP/123-QED}

\title{Studying power-grid synchronization with incremental refinement of model heterogeneity}
\author{B. Hartmann}%
\email{hartmann.balint@ek.hun-ren.hu}
\affiliation{Institute of Energy Security and Environmental Safety, HUN-REN Centre for Energy Research, P.O. Box 49, H-1525 Budapest, Hungary
}%
\author{G. Ódor}
\affiliation{Institute of Technical Physics and Materials Science, HUN-REN Centre for Energy Research, P.O. Box 49, H-1525 Budapest, Hungary
}
\author{K. Benedek}
\affiliation{Institute of Technical Physics and Materials Science, HUN-REN Centre for Energy Research, P.O. Box 49, H-1525 Budapest, Hungary
}
\affiliation{Department of Theoretical Physics, Budapest University of Technology and Economics, Budafoki út 8, H-1111 Budapest, Hungary}
\author{I. Papp}
\affiliation{Institute of Technical Physics and Materials Science, HUN-REN Centre for Energy Research, P.O. Box 49, H-1525 Budapest, Hungary
}%


\begin{abstract}
The dynamics of electric power systems are widely studied through the phase synchronization of oscillators, typically with the use of the Kuramoto equation. While there are numerous well-known order parameters to characterize these dynamics, shortcoming of these metrics are also recognized. To capture all transitions from phase disordered states over phase locking to fully synchronized systems, new metrics were proposed and demonstrated on homogeneous models. In this paper we aim to address a gap in the literature, namely, to examine how gradual improvement of power grid models affect the goodness of certain metrics. To study how the details of models are perceived by the different metrics, 12 variations of a power grid model were created, introducing varying level of heterogeneity through the coupling strength, the nodal powers and the moment of inertia. The grid models were compared using a second-order Kuramoto equation and adaptive Runge-Kutta solver, measuring the values of the phase, the frequency and the universal order parameters. Finally, frequency results of the models were compared to grid measurements. We found that the universal order parameter was able to capture more details of the grid models, especially in cases of decreasing moment of inertia. Even the most heterogeneous models showed notable synchronization, encouraging the use of such models. Finally, we show local frequency results related to the multi-peaks of static models, which implies that spatial heterogeneity can also induce such multi-peak behaviour.
\end{abstract}

\maketitle

\begin{quotation}
Modeling power-grid systems has got a major importance in present days as transformation to renewable energy sources requires the complete re-design of energy transmission.
Renewable energy sources can be located quite far from their consumption points because urban and industrial structures do not follow physical constraints and capabilities. Important examples are the sea coast vs
inland divisions in the case of wind power. Ill-constructed high-voltage (HV) power grids can cause catastrophic damages to economies as it was demonstrated in recent history via the emergence of large blackout events\cite{blackout1,blackout2,blackout3,blackout4,blackout5}. The probability distributions of such events was found to be fat-tailed, exhibiting power-law (PL) tails very often
~\cite{Car}. To understand them, self-organized critical direct current (DC) models have been constructed~\cite{car2} and have been shown to describe well the PL exponents of empirical values. However, many details could not be understood as power-grids work with alternating currents (AC) in which phase differences are the primary causes of the power-flows.
\end{quotation}

\section{\label{sec:level1}INTRODUCTION\protect\\}

AC modelling of power-grids have been proposed since the equivalence of swing equations to the second-order Kuramoto model was shown~\cite{fila}.
Failures leading to blackouts have been studied by composite Kuramoto and threshold models~\cite{Schäfer2018} and the PL tailed cascade failures could be modelled by them~\cite{POWcikk,Powfailcikk,USAEUPowcikk}. Network topological features, which lead to desynchronization by network fragmentation and Braess paradox phenomena, have been identified~\cite{Cohen1991, Witthaut2012, Witthaut2013, Fazlyaba2015, nagurney2016observation, Coletta2016, motter2018antagonistic,Olmi-Sch-19}. We have shown that these are basically consequences of quenched heterogeneity, which can be mitigated by the enhanced fluctuations, that arise naturally in the neighborhood of synchronization transition points, where power grids self-organize themselves by the competition of supply and demand~\cite{POWcikk,Powfailcikk,USAEUPowcikk,PhysRevResearch.6.013194,HARTMANN2024101491}. It was shown recently that heterogeneous networks represented by a mixture of inertial and non-inertial oscillators display synchronization transitions with varying exponent at the critical point~\cite{PARK2024115315}.

In power grid systems, the focus is often on the phase synchronization of the individual oscillators~\cite{kuraSchimanskiGeier} since their steady state is usually a stable limit cycle. To study the synchronization dynamics, several order parameters are used to characterize the dynamic state of the system. In the literature, there are numerous well-known Kuramoto order parameters, such as the complex order parameter~\cite{kuraSchimanskiGeier, STROGATZ20001}, the local order parameter measuring the phase coherence and its global variant~\cite{ARENAS200893}, a mean-field variant of the complex order parameter~\cite{BOCCALETTI2006175} or the one respecting network topology~\cite{gomez_loc_top_ord_param}.

Shortcomings of these metrics have been highlighted in a number of papers, most importantly by Ref. ~\cite{schroder2017}, who claim that existing order parameters are not fully suitable to characterize complex oscillator networks as they don't capture all transitions from incoherence over phase locking to full synchrony for arbitrary, finite networks. Hence a universal order parameter was also introduced, which captures partial phase locking, respects the topology of the network, and has been shown to increase monotonically with the coupling strength. It is worth noting that similar concepts, i.e. the use of composite indicators were also suggested by the engineering community to have a more coherent view on synchronization and stability phenomena~\cite{9613746,2022Taczi,10035478}.

In this paper, we aim to address a research gap in the literature, namely to examine how different modeling assumptions regarding the heterogeneities of a power grid are captured by the different order parameters. In the different scenarios, to analyze and compare the dynamic behavior of the various models, we will use the frequency spread, the global order parameter, and the newly proposed universal order parameter by ref.~\cite{schroder2017} as the main measures. We also present a frequency analysis of the simulation results and confirm q-Gaussian distributions, matching real data distributions presented in one of our earlier work~\cite{HARTMANN2024101491}.

The remainder of the paper is structured as follors. Section \ref{sec:2} introduces the synchronization model and the twelve grid models. Section \ref{sec:3} presents the results organized around five aspects. Finally, these results are discussed in Section \ref{sec:4}, and conclusions are drawn.

\section{Data and models\label{sec:2}}

\subsection{The synchronization model}

Modeling power-grid systems come in different flavors, but at the heart of most approaches describing the time evolution lies the so-called swing equations~\cite{swing}, set up for mechanical elements (e.g.~rotors in generators and motors) with inertia. Mathematically it is formally equivalent to the second-order Kuramoto equation~\cite{fila}, for a network of $N$ oscillators with phases $\theta_i(t)$.

To investigate the effects of different parametrizations and to facilitate benchmarking with previous results, we used a more specific form~\cite{Taher_2019,POWcikk,USAEUPowcikk,HARTMANN2024101491}, which includes dimensionless electrical parametrization and approximations for unknown ones:
\begin{equation}\label{kur2eq}
{\ddot{{\theta }}}_{i}+\alpha {\ }{\dot{{\theta}}}_{i}=\frac{P_i}{I_i\omega_S}
+\frac{{P}_{i}^{max}}{{I}_{i}{\ }{\omega }_{S}}{\ }\sum
_{j=1}^{N} {W}_{\mathit{ij}}{\ }\sin \left({\theta }_{j}-{\theta}_{i}\right) + \Omega_i \ .
\end{equation}
In this equation $\theta_i$ is the phase angle, $\omega_i=\dot{\theta_i}$, is the frequency of node $i$, $\alpha$ is the dissipation or damping factor, $W_{ij}$ is the coupling strength and $P_i$ is the source/load power. Furthermore, $I_i$ denotes the rotation inertia, $\omega_S$ the system frequency, and $P_i^{max}$ the maximal transmitted power in the system. Note that since the Kuramoto-equation is invariant to addition, the intrinsic frequency of nodes ($\Omega_i$=50 Hz in Europe) was omitted in the calculations. Our frequency results show the deviations from this value. 

If we know more details of the electrical parameters we can cast this into the form with real physical dimensions:
\begin{equation}\label{eq:kur2_phys}
    \dot{\omega}_i = -\frac{D_i \omega_i}{M_i \omega_S} + \frac{L_i}{M_i \omega_S} + \sum_{j=1}^{N} \frac{Y_{ij} V_i V_j}{M_i \omega_S} \sin(\theta_j - \theta_i), 
\end{equation}
where $D_i$ has dimension of $\left[\frac{\si{\kg\cdot\m^2}}{\si{\s}^2}\right]$ and describes the damping effect of element $i$ in the system, $L_i$ $\left[\frac{\si{\kg\cdot\m^2}}{\si{\s}^3}\right]$ is the power capacity of node $i$, $Y_{ij}=\frac{1}{X_{ij}}$ $\left[\frac{1}{\si{\ohm}}\right]$ is the susceptance of lines, the inverse of reactance, $V_i$ $[\si{\volt}]$ is the nodal voltage level and $M_i$ $\left[\si{\kg\cdot\m^2}\right]$ is the moment of inertia.

Topological heterogeneity of power grids is the result of two factors, (i) the structure and connectivity of the grid itself, and (ii) the heterogeneity of power line capacities and nodal behaviors, as it was presented in our recent work~\cite{HARTMANN2024101491}. In the second-order Kuramoto equation, these varying properties are represented by the parameters $L_i$, $Y_{ij}$ and $M_i$.

The time step resolution of the calculations was set to be $\Delta t= 0.25\;\si{\s}$ and $\alpha=0.1$ was used, similarly as in Refs~\cite{olmi, menck}. Ref.~\cite{menck} also used  $\alpha=0.4$. In this way, the results of the Kuramoto equation become dimensionless.

To model station fluctuations, we have added a multiplicative, quenched noise to the equations of motion (\ref{kur2eq}),(\ref{eq:kur2_phys}) as additional source/sink terms
\begin{equation}\label{eq:noise}
\eta_{i,j} = 0.05 \xi_j \frac{D_i\omega_i}{M_i \omega_s} ,
\end{equation}
where $\xi_j \in N(0.1)$ is drawn from a zero centered Gaussian distribution. To solve the equations of motion we used an adaptive Runge-Kutta-Fehlberg method~\cite{watts_shampine_burkardt_r8_rkf45_2004} from the package Numerical Recipies~\cite{Press2007}.

We investigated the standard synchronization measures of the phases $R(t)$ and the frequency spread $\Omega(t)$, called the frequency order parameter. We measured the Kuramoto phase order parameter:
\begin{equation}\label{ordp}
z(t_k) = r(t_k) \exp\left[i \theta(t_k)\right] = 1 / N \sum_j \exp\left[i \theta_j(t_k)\right] \ .
\end{equation}
Sample averages over different initial fluctuations for the phases
\begin{equation}\label{KOP}
R(t_k) = \langle r(t_k)\rangle
\end{equation}
and for the variance of the frequencies
\begin{equation}\label{FOP}
        \Omega(t_k) = \frac{1}{N} \langle \sum_{j=1}^N (\overline\omega(t_k)-\omega_jt_k))^2 \rangle
\end{equation}
were determined, where $\overline\omega(t_k)$ denotes the mean frequency within each respective sample at time step
$t_k = 1 + 1.08^{k}$, $k=1,2,3...$. 
Sample averages were computed from the solutions with hundreds of independent self-frequency realizations (i.e. $\eta_{i,j}$) for each control parameter.

Besides, we measured a more complex order parameter suggested for the second-order Kuramoto model, which claimed to accurately track the degree of partial phase locking and synchronization\cite{schroder2017}
\begin{equation}
    r_{iui}(t_k) = 1/(\sum_{i,j}^N w_{ij}) \sum_{i,j}^N w_{ij} \cos(\theta_i-\theta_j) 
\end{equation}
and it's sample and temporal average in the steady state:
\begin{equation}\label{eq:runi}
    R_{uni}=\langle r_{iui}(t_k) \rangle
\end{equation}

The fluctuations of the order parameters are measured by the standard deviations of the sample and temporal averages in the steady state, typically after 250 s transient time.

\subsection{\label{sec:level2}The grid model}
We chose the Hungarian high-voltage (\SI{132}, \SI{220}, and \SI{400}{kV}) network to create our grid models. 
The network consists of 387 nodes and 640 edges, and its most important features are presented in Table \ref{tab:net_char}. 
Cross-border transmission lines were reduced to their domestic terminals as sources or consumers, 
thus resulting in a standalone synchronous system. In the modeled loading state, 351 nodes behave as consumers and 36 as sources.

\begin{table}[!ht]
\caption{Selected characteristics of the Hungarian power grid. $N$ and $E$ denote the number of nodes and edges.$\langle k\rangle$ is the average degree, $L$ is the average shortest path, $C$ is the clustering coeffient, $Q$ is the modularity quotient  \cite{Newman2006-bw}. $\sigma$ and $\omega$ are small-world metrics according to Fronczak et al.\cite{Fron} and Telesford et al.\cite{doi:10.1089/brain.2011.0038}, $Eff$ is the global efficiency of the network\cite{latora2001}, and  $\gamma$ is the decay exponent of the exponential of the degree distribution\cite{sole2008}.}
\centering 
\begin{tabular}{|c|c|c|c|c|c|c|c|c|c|} 
\hline 
$N$ & $E$ & $\langle k \rangle$ & $L$ & $C$ & $Q$ & $\sigma$ & $\omega$ & $Eff$ & $\gamma$ \\ 
\hline 
387 & 640 & 3.307 & 6.566 & 0.077 & 0.4666 & 6.855 & 0.521 & 0.177 & 1.726 \\
\hline 
\end{tabular}
\label{tab:net_char}
\end{table} 

In order to study how modeling depth is perceived by the different order parameters, we created 12 different variations of the Hungarian network. These variations introduce heterogeneity to the parameters $W_{ij}$, $L_i$,  and $M_i$. The resulting representations thus range from completely homogeneous networks, which are the most widely covered in related literature, to completely heterogeneous ones, where electric parameters and nodal behaviors are defined using the actual data and measurements of the Hungarian system.
The following assumptions are used for the three parameters.

\begin{figure}[H]
    \centering
    \includegraphics[width=85mm]{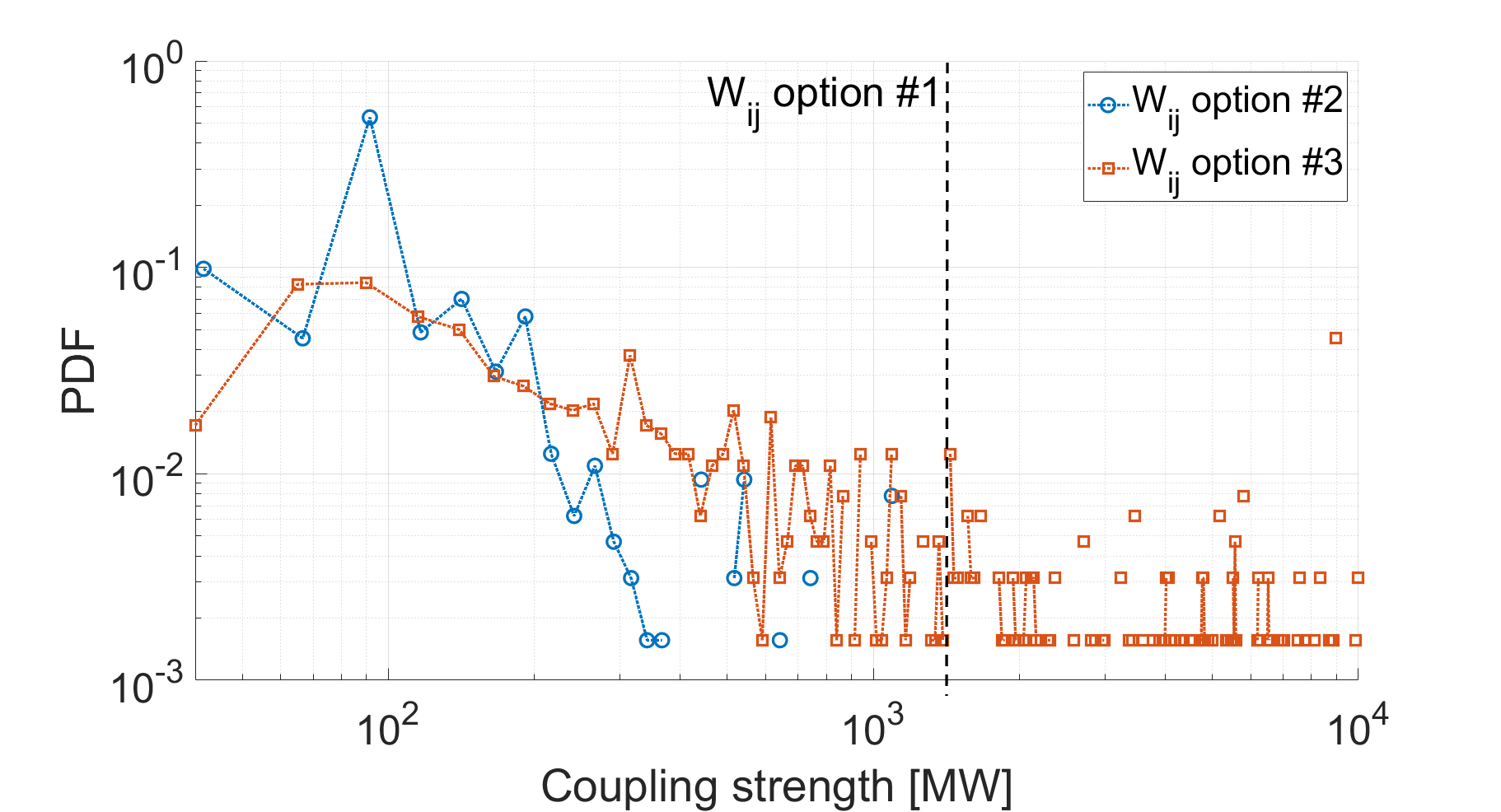}
    \caption{PDFs of the various options for the coupling strength. Note that option 2 represents less heterogeneous but lower values, while 3 is more heterogeneous with higher values; their respective medians are $\approx100$ and $\approx400$ MW.}
    \label{fig:W_ij_PDF}
\end{figure}

\begin{figure}[H]
    \centering
    \includegraphics[width=85mm]{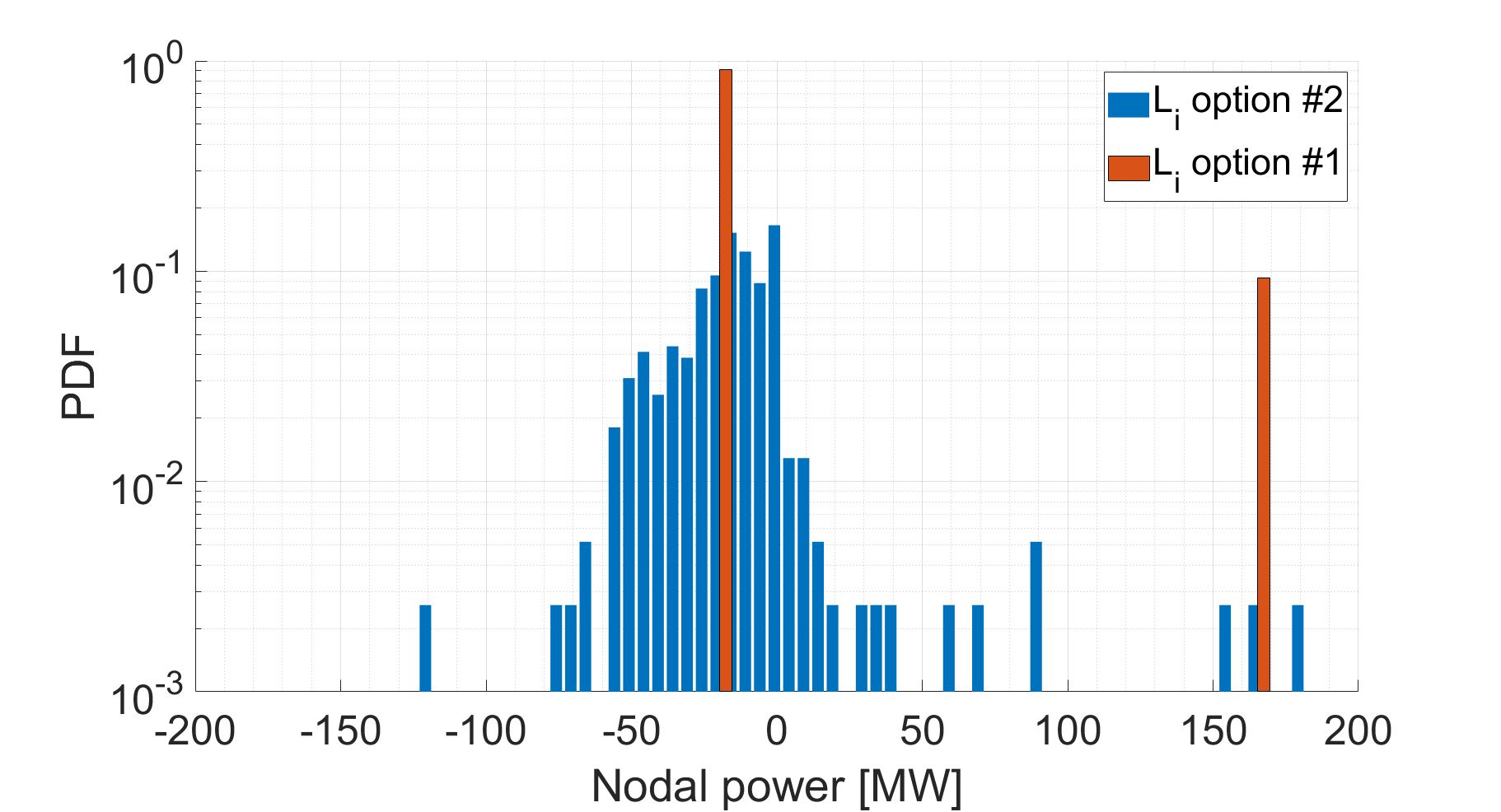}
    \caption{PDFs of the various options for the nodal power.}
    \label{fig:L_i_PDF}
\end{figure}

\begin{itemize}
    \item $W_{ij}$, coupling strength (Fig. \ref{fig:W_ij_PDF}):
        \begin{enumerate}
            \item Identical value for each edge, the value corresponding to the largest thermal capacity (ampacity) limit in the system (approx. 1400 MW). This option represents the benchmark used by e.g. Refs.~\cite{olmi, menck}.
            \item Unique value for each edge, depending on their actual thermal capacity limits (range between 40 and 1400 MW).
            \item Unique value for each edge, depending on their actual admittance $Y_{ij}$ and voltage level.
        \end{enumerate}
    \item $L_i$, nodal power (Fig. \ref{fig:L_i_PDF}):
        \begin{enumerate}
            \item The sources ($L_i<0$) are distributed equally among the nodes representing power plants and the consumers ($L_i>0$) are distributed equally among nodes truly representing consumption. This is a slightly modified assumption of Ref. \cite{menck}, where half the nodes correspond to consumers ($L_i>0$), while the other half to power sources ($L_i<0$).
            \item Every $L_i$ value is uniquely assigned, based on measured data (SCADA).
        \end{enumerate}
    \item $M_i$, moment of inertia:
        \begin{enumerate}
            \item Aligning with the literature, we set a constant value of 40,000 $kg\cdot m^2$ for $M_i$, corresponding to a $\SI{400} {\mega\watt}$ gas turbine power plant as in~\cite{menck, olmi}. 
            \item We evenly distribute the moment of inertia among $\SI{400}{\kilo\volt}$ and $\SI{220}{\kilo\volt}$ nodes, which host the majority of synchronous generators (conventional power plants). $M_i$ = 16,665 $kg\cdot m^2$.
            \item We evenly distribute the moment of inertia along all nodes of the model. $M_i$ = 1593 $kg\cdot m^2$.
            \item We set unique values based on measured data and whether the node actually hosts a synchronous machine or not. Median of $M_i$ values is 1382 $kg\cdot m^2$.
        \end{enumerate}
\end{itemize}

In the following the 12 scenarios will refer to the different models as shown in Table \ref{tab:models}.

\begin{table}[H]
\caption{The different scenarios, representing increasing level of heterogeneity in modelling. 
The numbers from the table correspond to the list number of the assumption for the respective parameter.}
\centering
\begin{tabular}{c|cccccccccccc}
\textbf{Scenario \#} &1 &2 &3 &4 &5 &6 &7 &8 &9 &10 &11 &12 \\
\hline
\textbf{$W_{ij}$} &1 &1 &1 &1 &2 &2 &2 &2 &3 &3 &3 &3 \\
\textbf{$L_i$} &1 &1 &1 &2 &1 &1 &1 &2 &1 &1 &1 &2 \\
\textbf{$M_i$} &1 &2 &3 &4 &1 &2 &3 &4 &1 &2 &3 &4\\
\hline
\end{tabular}
\label{tab:models}
\end{table}

\section{Results\label{sec:3}}

In the following, results of the synchronization studies are presented in a sequential, interdependent way. First, the coupling strength was varied to compare how different order parameters display criticality (Section \ref{sec:3A}). Then the transient behaviour of the three order parameters, $R$, $\Omega$ and $R_{uni}$ are analysed in sections \ref{sec:3B}, \ref{sec:3C} and \ref{sec:3D}, respectively. Section \ref{sec:3E} compares the frequency data of the simulations to grid frequency data presented in Ref. \cite{HARTMANN2024101491}.

Note that to assist the interpretation of the results, not all scenarios are displayed in all figures.
\subsection{\label{sec:3A}Dependence of the critical point on the models}

We assume that power-grid-like systems operate near the state of self-organized criticality (SOC)~\cite{SOC,dobson2007c,Schmietendorf_2014,POWcikk}. This means that the system is not operated on $100\%$ load capacity, but usually at a lower level. To mimic the not fully loaded behavior we can cast equation \eqref{kur2eq} in the following form:
\begin{equation}\label{kur2eq2}
    \dot{\omega}_i = -\frac{D_i \omega_i}{M_i \omega_S} + \frac{L_i}{M_i \omega_S} + \lambda \sum_{j=1}^{N} \frac{Y_{ij} V_i V_j}{M_i \omega_S} \sin(\theta_j - \theta_i) + \eta_{i,j}, 
\end{equation}

The multiplicative factor $\lambda$ in front of the interaction term is the chosen constant for the system and its value corresponds to different load levels. In a mathematical sense this maps to a changing coupling strength, which allows us to identify the SOC behavior by analyzing the standard deviation of the Kuramoto order parameter.

\begin{figure}[H]
    \centering
    \includegraphics[width=85mm]{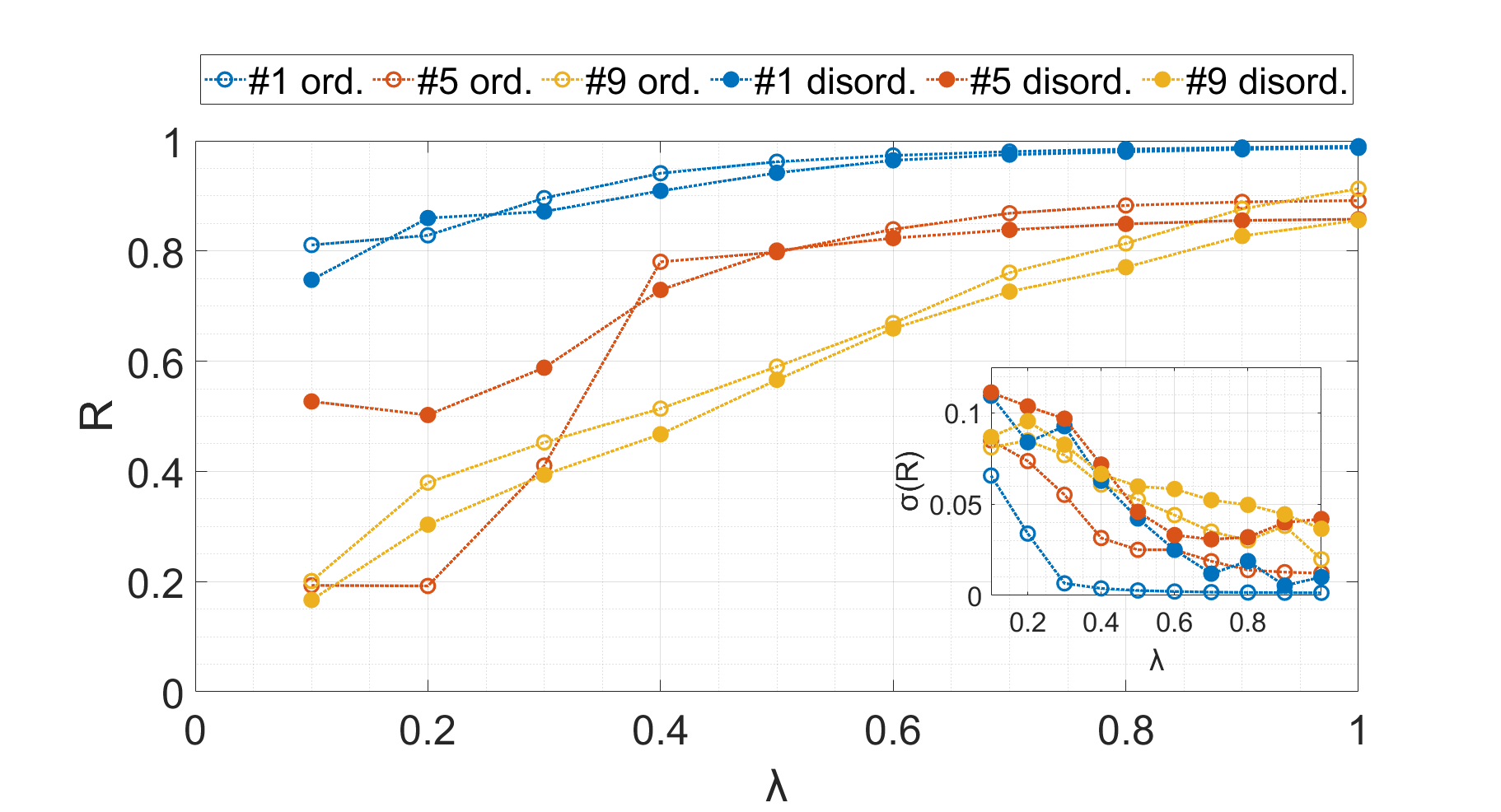}
    \caption{Main figure: the Kuramoto order parameter $R$ in the steady state, defined by Eq.~\eqref{KOP} as the function of different $\lambda$ values for scenarios $1$, $5$ and $9$. Inset: the corresponding standard deviations of the Kuramoto order parameter, $\sigma(\mathrm{R})$.}
    \label{fig:R_lambda}
\end{figure}

To identify the cross-over point to synchronization (Fig. \ref{fig:R_lambda}), we varied the $\lambda$ parameter from $0.1$ to $1.0$ with steps of $0.1$. For the sake of completeness, we performed analysis starting the system from a phase-ordered state, i.e. all oscillators have the same initial phase and some noise, or from a disordered phase, i.e. all the oscillators have random initial phase assignment. In the figures of this section, unfilled markers will represent the phase-ordered states, and filled markers are the disordered states.

\begin{figure}[H]
    \centering
    \includegraphics[width=85mm]{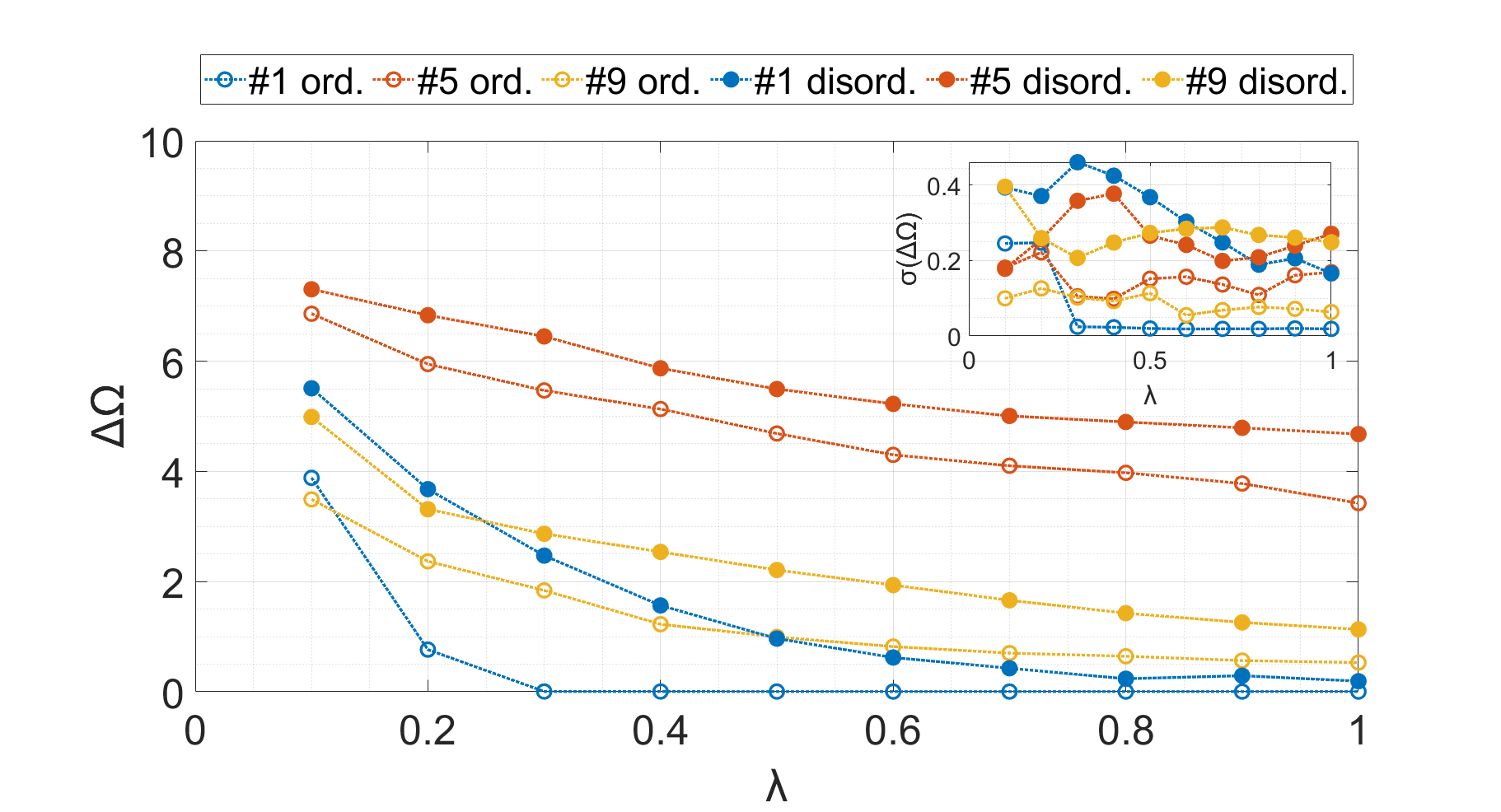}
    \caption{Main figure: Frequency spread measured in the steady state, defined by Eq.~\eqref{FOP}, across various $\lambda$ values. Note that $\Delta\Omega$ means the deviation compared to the nominal $\Omega_i$. Inset: standard deviation of the frequency spread.}
    \label{fig:W_lambda}
\end{figure}

\begin{figure}[H]
    \centering
    \includegraphics[width=80mm]{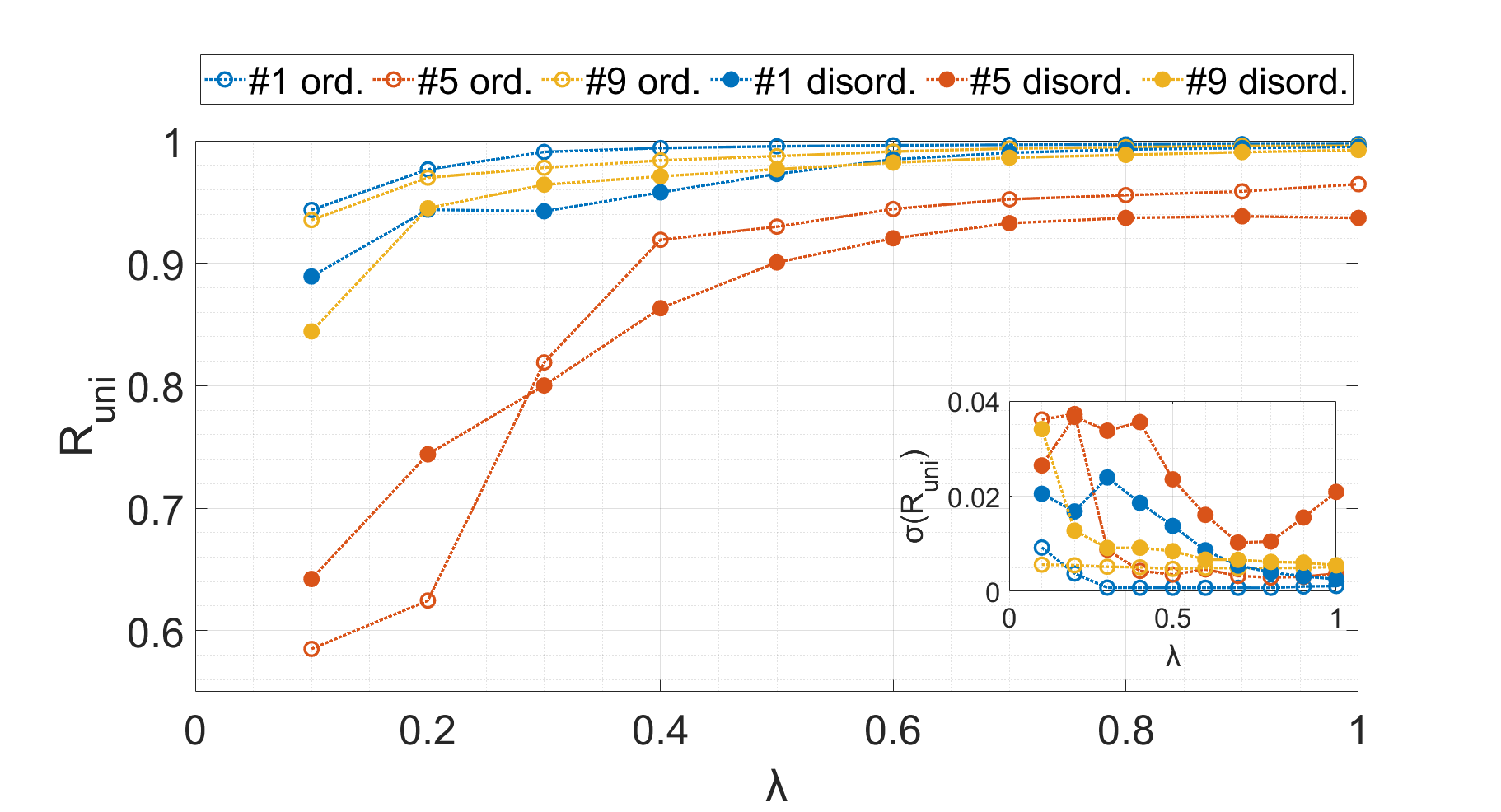}
    \caption{Main figure: the universal order parameter, $R_\mathrm{uni.}$ in the steady state, defined by Eq.~\eqref{eq:runi}. Inset: standard deviation of $R_\mathrm{uni.}$ as the function of different $\lambda$ values.}
    \label{fig:R_uni_lambda}
\end{figure}

We have chosen model scenarios $1$, $5$, and $9$ for finding the optimal $\lambda$ value, as these scenarios represent homogeneous nodal behavior with three different options for defining the coupling strength. As it was shown in Fig. \ref{fig:W_ij_PDF}, the PDFs of these options have different spreads and medians as well, as they are based on different technical assumptions. We performed the statistical calculation on a minimum $1920$ sample and a maximum $5000$ for each $\lambda$ value. The results are shown in Figs. \ref{fig:R_lambda}-\ref{fig:R_uni_lambda} for $R$, $\Omega$ and $R_{uni}$, respectively.

Fig. \ref{fig:R_lambda} shows that phase-ordered and phase-disordered initial conditions show little to no difference if $\lambda>0.4$; below this threshold scenario 5 indicates non-trivial behavior. It is worth noting that over the majority of the range, scenario 5 reaches higher values than scenario 9, despite representing lower (but more homogeneous) coupling strengths.

Different results were obtained for $\Omega$-s, as shown in Fig. \ref{fig:W_lambda}. Considering the deviations of the frequency, scenario 5 behaves distinctively different than the other two, which might be directly related to having the lowest coupling strength.

Fig. \ref{fig:R_uni_lambda} displays that the $R_{uni}$ order parameter practically increases monotonically as expected. Standard deviation of scenario 1 for phase-disordered results shows a peak at $\lambda=0.3$, while scenario 5 displays multiple local peaks.
Finally, it is important to notice that $R_{uni}$ captures high synchronization for scenarios 1 and 9, regardless of the value of $\lambda$.

\subsection{\label{sec:3B}Kuramoto order parameter transient behavior}

Fig. \ref{fig:R_therm} in the Appendices shows the Kuramoto order parameter, $R$ in the transient from different initial conditions. It can be seen that the highest $R$ values are reached by scenarios 1-4. In these scenarios, the coupling strength, $W_{ij}$ is homogeneous and high, allowing quick synchronization even for scenarios 4 with highly heterogeneous nodal behavior. Starting from ordered initial conditions, both pairs of scenarios 5-6 and 9-10 reach similar $R$ values ($\approx0.84$ and 0.7, respectively). This suggests that solely the decrease of the moment of inertia, $M_i$ does not affect the order parameter. Slightly different observations are seen if the simulations are started from phase disordered conditions, where there are visible differences between scenarios 5 and 6, and 9 and 10. Finally, It can also be seen that increasing the heterogeneity of the moment of inertia, $M_i$, decreases the value of the order parameter, as expected.

\subsection{\label{sec:3C}The $\Omega$ order parameter transient}

Fig.~\ref{fig:W_therm} in the Appendices shows the evolution of the order parameter $\Omega$. Compared to the curves of $R$, bigger differences can be noticed. Scenario 1 and 2 in case of phase ordered initial conditions show very low frequency spread ($\Omega\approx10^{-5} - 10^{-4}$). For all other cases, $\Omega$ values are between $\approx10^{0} - 10^{5}$. The highest spreads are seen in case of scenarios 4, 8 and 12 (the most heterogeneous models), followed closely by scenarios 3, 7 and 11 (small, but homogeneous $M_i$s). It can also be seen that for these six scenarios, changing  coupling strengths, $W_{ij}$ does not affect the steady-state values of $\Omega$ too much if all other parameters are fixed, thus heterogeneity in the edge behavior seems to be irrelevant.

\subsection{\label{sec:3D}The new order parameter $R_{uni}$}

Fig. \ref{fig:R_uni_therm} in the Appendices shows the evolution of $R_{uni}$ from different initial conditions. Compared to $R$ the most important difference is that significantly higher order parameter values are reached for all scenarios. Starting from ordered initial conditions, Nine of twelve scenarios reach at least $R=0.9$, with the exceptions being scenarios 7, 8 and 12. Similarly to the other order parameters, the highest synchronization is shown by the highly homogeneous models of scenario 1 and 2, but $R_{uni}$ ranks scenarios 9 and 10 almost equally high, regardless of the initial conditions. This suggests that using unique, admittance-based values for the coupling strength does not jeopardize stability, despite their wide distribution (see $W_{ij}$ option 3 in Fig. \ref{fig:W_ij_PDF}).

If only the heterogeneity of the coupling strength, $W_{ij}$ is changed, better synchronization is found in the case of the scenarios, where the values of $W_{ij}$ are calculated from admittances (scenarios 9-12) as compared to the ones, where actual thermal capacity was considered (5-8). The technical reason behind this phenomena is possibly that admittances are primarily affected by the conductors themselves, thermal limitations are indirectly affected by protection settings as well. (E.g. in case of overloading it is more likely that the sag of the power line at critical spans between two towers will exceed life safety barriers.

\begin{figure}[H]
    \centering
    \includegraphics[width=85mm]{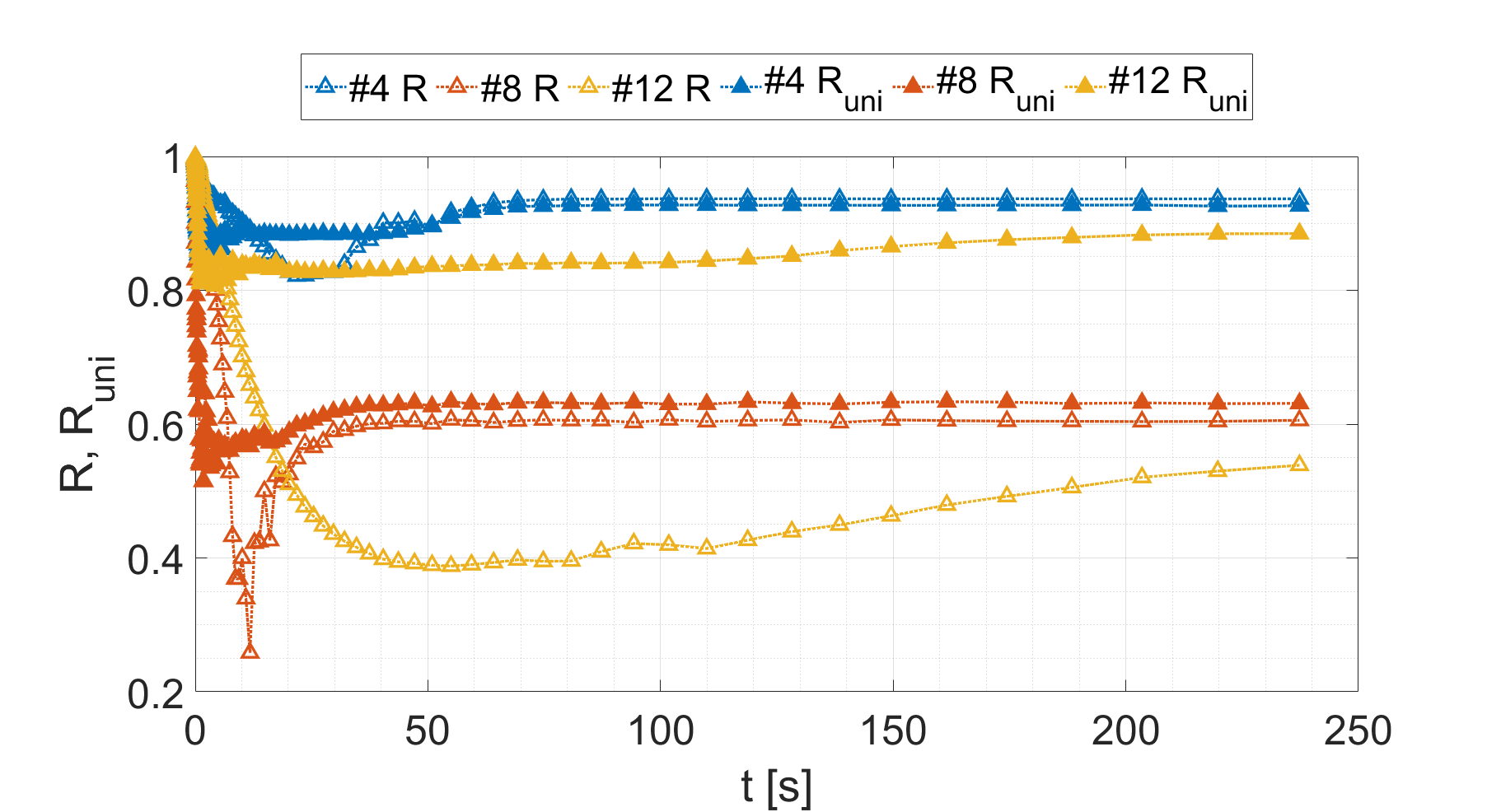}
    \caption{Evolution of $R$ and $R_{uni}$ for phase ordered initial conditions at $\lambda=0.5$.}
    \label{fig:R_vs_R_uni_4-8-12_ord}
\end{figure}

\begin{figure}[H]
    \centering
    \includegraphics[width=85mm]{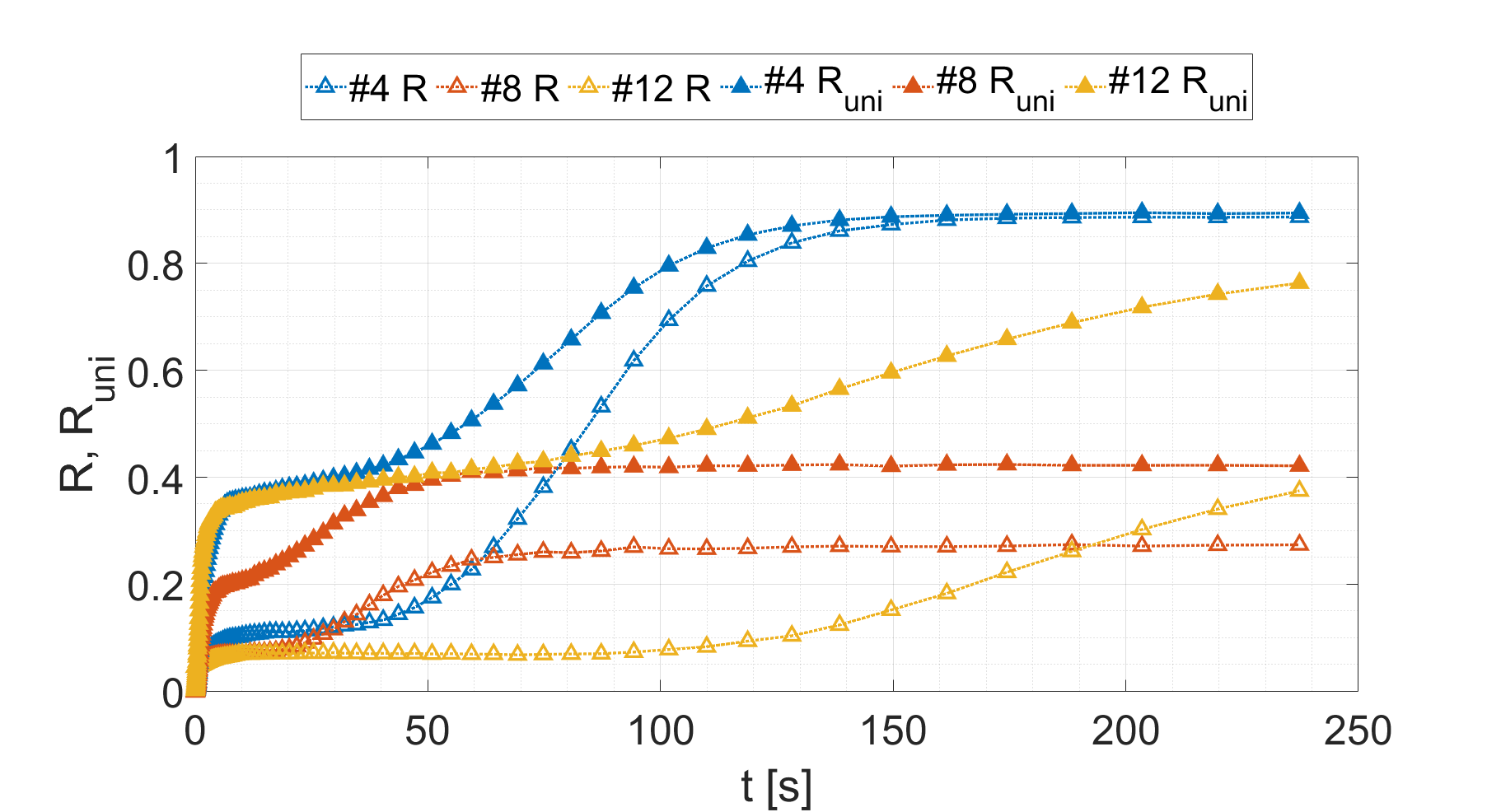}
    \caption{Evolution of $R$ and $R_{uni}$ for phase disordered initial conditions at $\lambda=0.5$.}
    \label{fig:R_vs_R_uni_4-8-12_disord}
\end{figure}

If we compare curves with the same coupling strength (shown with same colors), it is seen that the benchmark of literature, namely assuming the moment of inertia of a large power plant at each node, is a significant factor in reaching high synchronization. These models are demonstrably overly optimistic, especially with the increase of non-synchronous generation in the mix~\cite{Hartmann2019}. In Figs. \ref{fig:R_vs_R_uni_4-8-12_ord} and \ref{fig:R_vs_R_uni_4-8-12_disord} we show that more realistic modeling of the coupling strength (and the moment of inertia) does not inhibit synchronization of $R_{uni}$, thus its use over $R$ should be favored, especially when using the Kuramoto model for grid case studies. These findings are seen both for ordered and disordered initial conditions.

Finally, when setting unique coupling strength for each edge depending on their admittance, the values slowly increase over time, eventually reaching similar levels of synchronization as more homogeneous models for $R_{uni}$. However, if strength is set using capacity limits, a saturation is seen, and high synchronization levels are not expected to be reached.

\subsection{\label{sec:3E}Frequency distributions}

Testing the predictive power of the Kuramoto-based modeling is a great challenge and has not been done on the quantitative level on realistic power-grids, according to our knowledge. Here we show node frequency results obtained by different levels of parameter approximations. We calculated the PDF-s at nodes obtained in the steady state form samples at the last 10 time steps and from thousands of independent realizations, which differ in the input/output power by the addition of the small quenched fluctuation values as shown in Eq.(\ref{eq:noise}).

\begin{figure}[H]
    \centering
    \includegraphics[width=80mm]{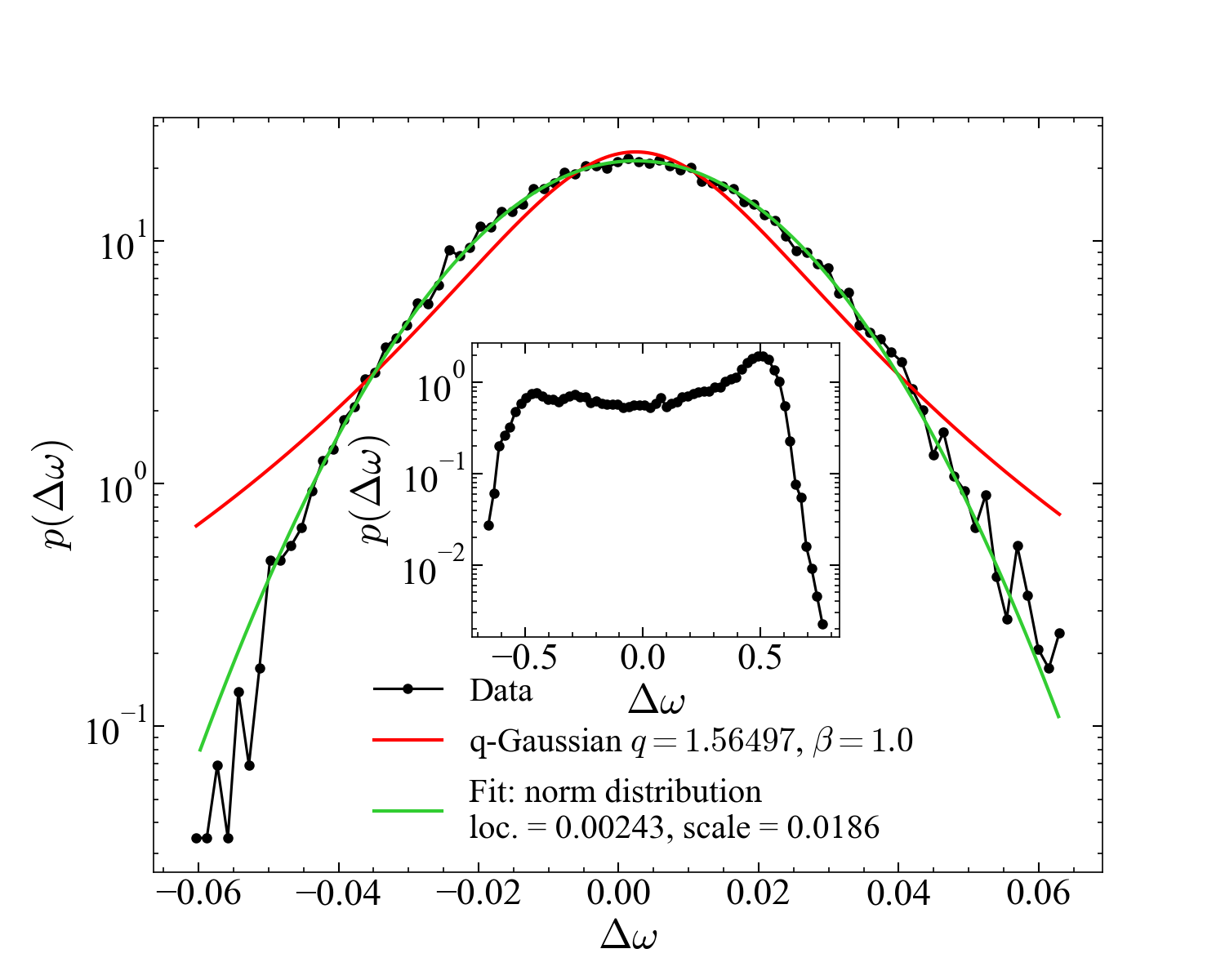}
    \caption{Local frequency fluctuation distributions with respect to $\Omega_i$, obtained for Scenario 9, corresponding to DETK substation for $\lambda=0.5$ with different numerical fits. The best fit is obtained with the normal distribution. The q-Gaussian provides $q \simeq 1.565$, close to the measurements. This particular node was chosen, because real data was fitted for the same substation in one of our previous works \cite{HARTMANN2024101491}. Similar distribution results were obtained for the other substations (e.g. Békéscsaba, Győr) as well. Multi-peak behavior of the frequency distribution at the same node, possible in the case of $\lambda\neq\lambda_c$ for $\lambda=0.3$.}
    \label{fig:freq-dist}
\end{figure}

These calculations were done for each scenario and for each $\lambda$.
We tried to fit the PDF-s with the 8 most popular distributions: Gaussian, exponential, Student's t, log-normal, Pareto, double Weibull, generalized extreme value, and beta, from the Python distfit~\cite{Taskesen_distfit_is_a_2020} package as well as by the q-Gaussian functions as this distribution was commonly fitted well other HV studies of AC electrical data~\cite{Sch_fer_2016,HARTMANN2024101491}. Agreement with the q-Gaussian is remarkably good for lower $\lambda$ values, see Fig~\ref{fig:freq-dist} in case of
Scenario 9.

For $\lambda > \lambda_c$ we found multi-peak behavior as displayed on the inset of Fig.~\ref{fig:freq-dist} even though the width of the frequency spread decreases by increasing $\lambda$. Multi-peak frequency behavior is very common in the European power grid, especially in islands, like Great Britain, Ireland, and Mallorca.~\cite{Kruse_2020,Kraljic_2023}. A numerical analysis based on the extension of the swing equations with a time-dependent damping factor could reproduce such global frequency fluctuations, suggesting that the system is wandering around the nominal 50Hz peak~\cite{Kraljic_2023}.

Multi-peak distribution poses a challenge for current power grid control systems, which typically have a three-step mechanism~\cite{BEVRANI2021107114}. The situation even deteriorates more with the introduction of renewable power sources or power electronics-based distributed generators~\cite{BEVRANI2021107114} as this leads to the reduction of the rotational inertia in a power system which can negatively affect the grid frequency response. This can induce skewed and/or multi-peak frequency distributions~\cite{8973560,9805854} to which conventional frequency control strategies can react poorly~\cite{wolff2019heterogeneities,10545571}. This also poses a new challenge to the scientific and engineering community in developing and testing new control strategies in electric grids.

Our swing equation solutions on the full Hungarian power-grid network suggest that even with constant parameters multi-peak frequency behavior can emerge when we overload the system with global power transmission above the synchronization point $\lambda_c$.
Thus, beyond temporally different behavior we can also find sub-peaks on static power grids, due to the network heterogeneity. Note, that in our previous large-scale simulations, we showed different synchronization behaviors at fixed control parameters in different communities of Europe~\cite{Deng_2024,HARTMANN2024101491}.
This may hint at the dangers in power grids with multi-peaks being out of optimal operation control even if the frequency spread is narrow.

For scenarios other than 9 fat-tailed distributions were also obtained. As they are very numerous publication of them will be published elsewhere.

\section{Discussion and conclusions\label{sec:4}}

As it was shown in the comparative analysis of Section \ref{sec:3}, the use of $R_{uni}$, proposed by Ref.\cite{schroder2017} is encouraged to display the differences of heterogeneous power grid models. We found that in contrast to $R$ and $\Omega$, this order parameter is able to capture the difference of the coupling strength, showing higher synchronization values when $W_{ij}$ is calculated based on the admittances of power lines. We also found that decreasing inertia of the system is more distinctly presented across the different scenarios, but does not inhibit synchronization. This feature of $R_{uni}$ is especially advantageous for case studies, as the benchmark models of the literature unnecessarily tend to overestimate the amount of inertia in the system.

Considering the incremental refinement of model heterogeneity we found that while completely heterogeneous models with unique nodal behavior based on SCADA measurements show lower synchronization in the early phases of the transients. However, when setting unique coupling strength for each edge depending on their admittance, $R_{uni}$ values slowly increase over time, eventually reaching similar levels of synchronization as more homogeneous models. In contrast, if unique coupling strength for each edge is set using thermal capacity limits, transients plateau after short time, showing lower order. These findings support and encourage the use of heterogeneous models, which previously might have been omitted due to relying on the traditional order parameter. As parts of the tested network can be interpreted as non-identical (and thus not symmetric) parts, the presence of synchronicity in steady-state, similar conclusions can be drawn as by Molnár, Nishikawa and Motter in Ref.~\cite{molnar2020network}.

The universal order parameter $R_{uni}$ used in the paper is more closely related to the power flows of the system, a complementary description to the Kuramoto order parameter $R$, which is related to the phases (and to the frequencies in case of the second-order Kuramoto-model). The underlying reasons of $R$ not capturing all model details could be (i) the ignorance of power losses and (ii) the ignorance of reactive power. To bridge these gaps, promising approaches are presented in the literature, where the voltage magnitudes ~\cite{Schmietendorf_2014,Taher_2019,PRXEnergy.1.013008,arinushkin2022influence} or the power losses~\cite{10004698,Soleimani24} are incorporated in the Kuramoto model.

We provided local frequency results for the whole Hungarian power-grids, agreeing quite well with empirical measurements~\cite{HARTMANN2024101491}. The calculated PDF-s of $\Delta{\Omega}$, with respect to the nominal 50 Hz exhibit similar width and shapes as those recorded and
published in~\cite{HARTMANN2024101491}. For $\lambda$-s, which drive the system above the synchronization point earlier we observed community dependent synchronization~\cite{Deng_2024,HARTMANN2024101491}.These are related to the frequency multi-peaks of static models we report now. This means, that
they occur not only in systems with time dependent parameters, but spatial heterogeneity can also induce them and warn for over-driven power grids, in the sense that they are away from the optimal SOC behavior.

\begin{acknowledgments}
Bálint Hartmann acknowledges the support of the Bolyai János Research Scholarship of the Hungarian Academy of Sciences (BO/131/23). Support from the Hungarian National Research, Development and Innovation Office NKFIH (K146736) is also acknowledged.
\end{acknowledgments}

\section*{Data Availability Statement}
The data that support the findings of
this study are available from the
corresponding author upon reasonable
request.

\section*{References}
\bibliography{aipsamp}

\section*{Appendices}
\begin{figure}[H]
    \centering
    \includegraphics[width=85mm]{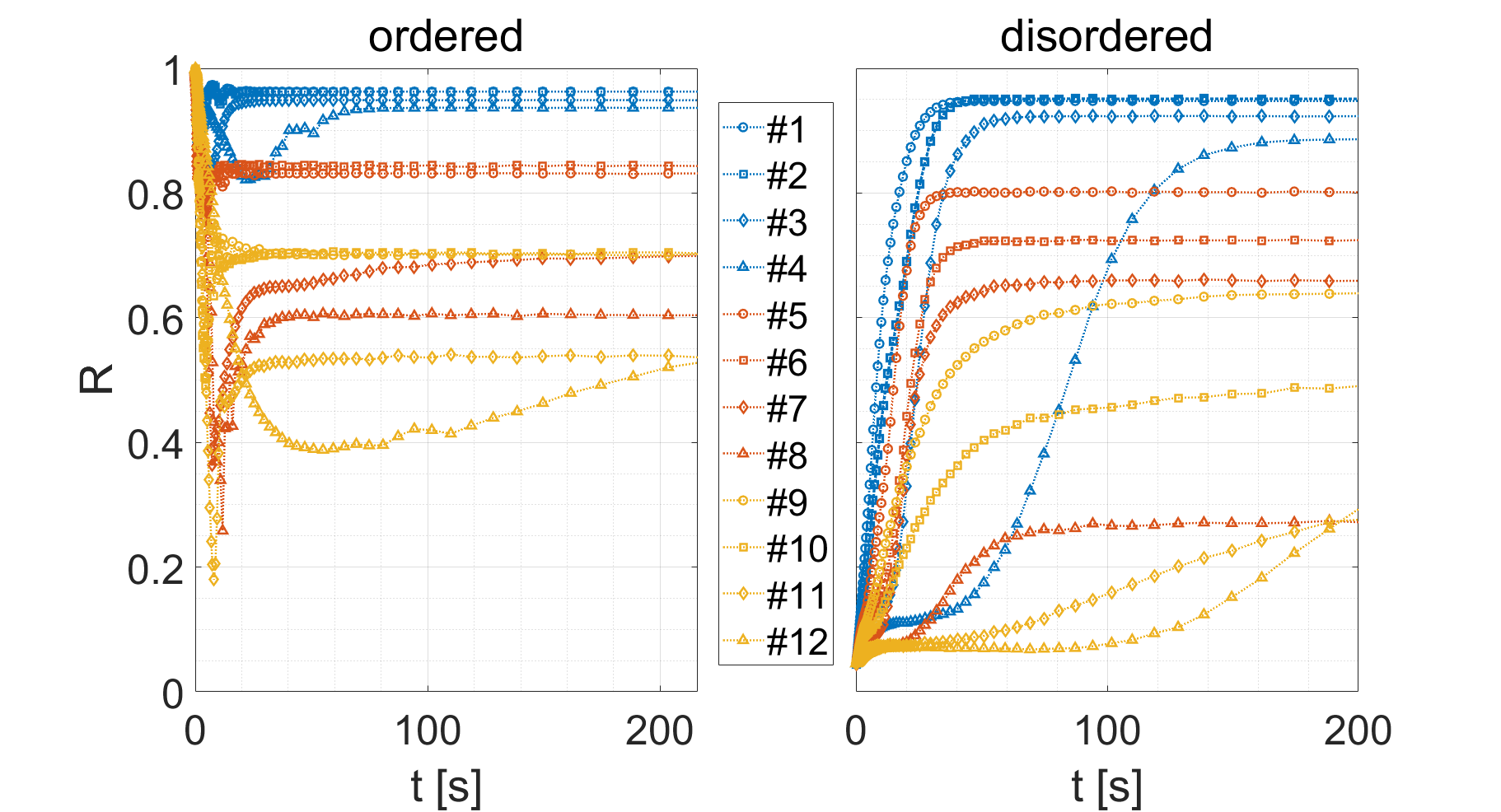}
    \caption{Kuramoto order parameter, $R$ for phase ordered and disordered initial conditions at $\lambda=0.5$.}
    \label{fig:R_therm}
\end{figure}

\begin{figure}[H]
    \centering
    \includegraphics[width=85mm]{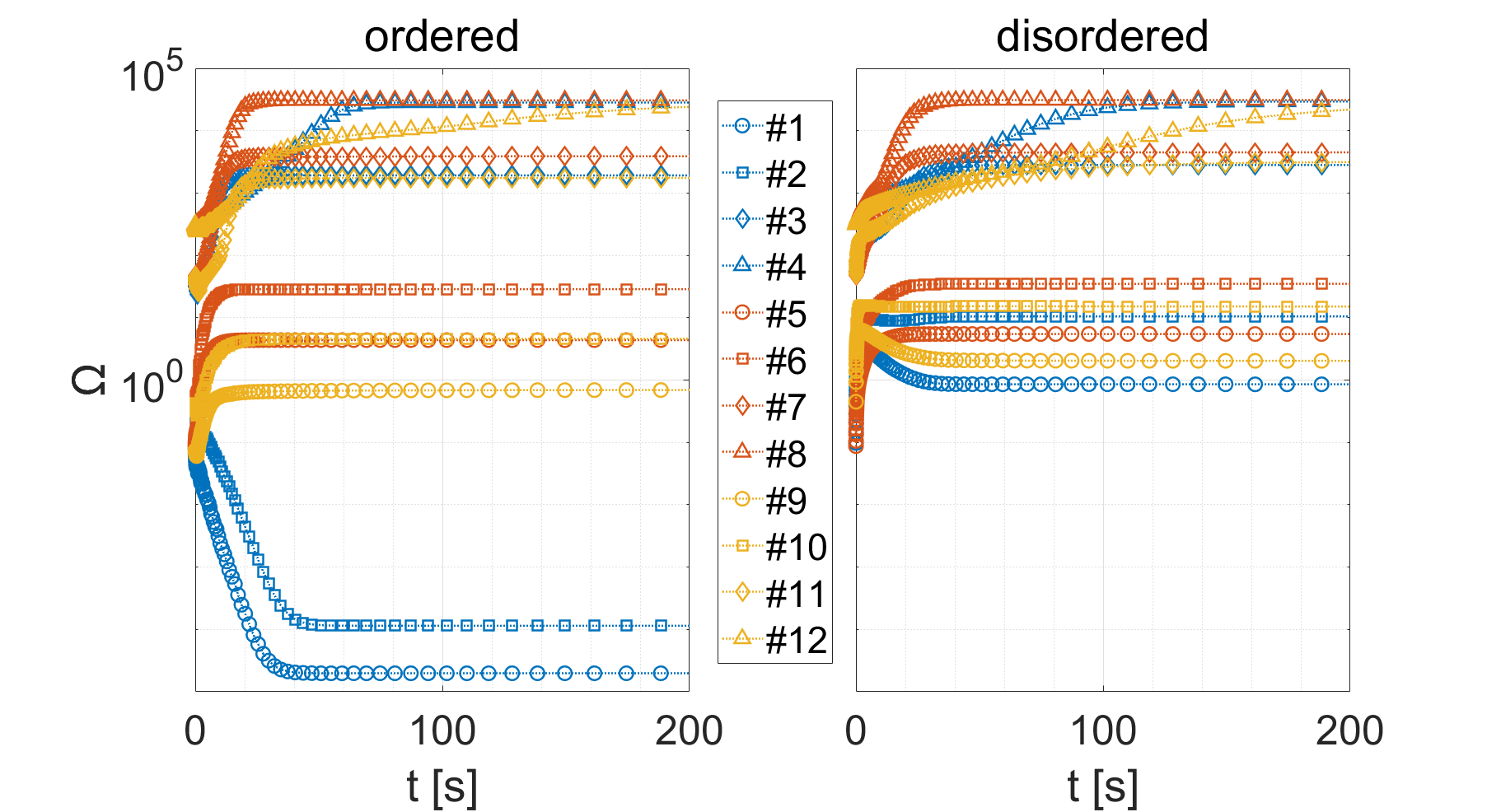}
    \caption{Evolution of frequency spread $\Omega$ for phase ordered and disordered initial conditions at $\lambda=0.5$.}
    \label{fig:W_therm}
\end{figure}

\begin{figure}[H]
    \centering
    \includegraphics[width=85mm]{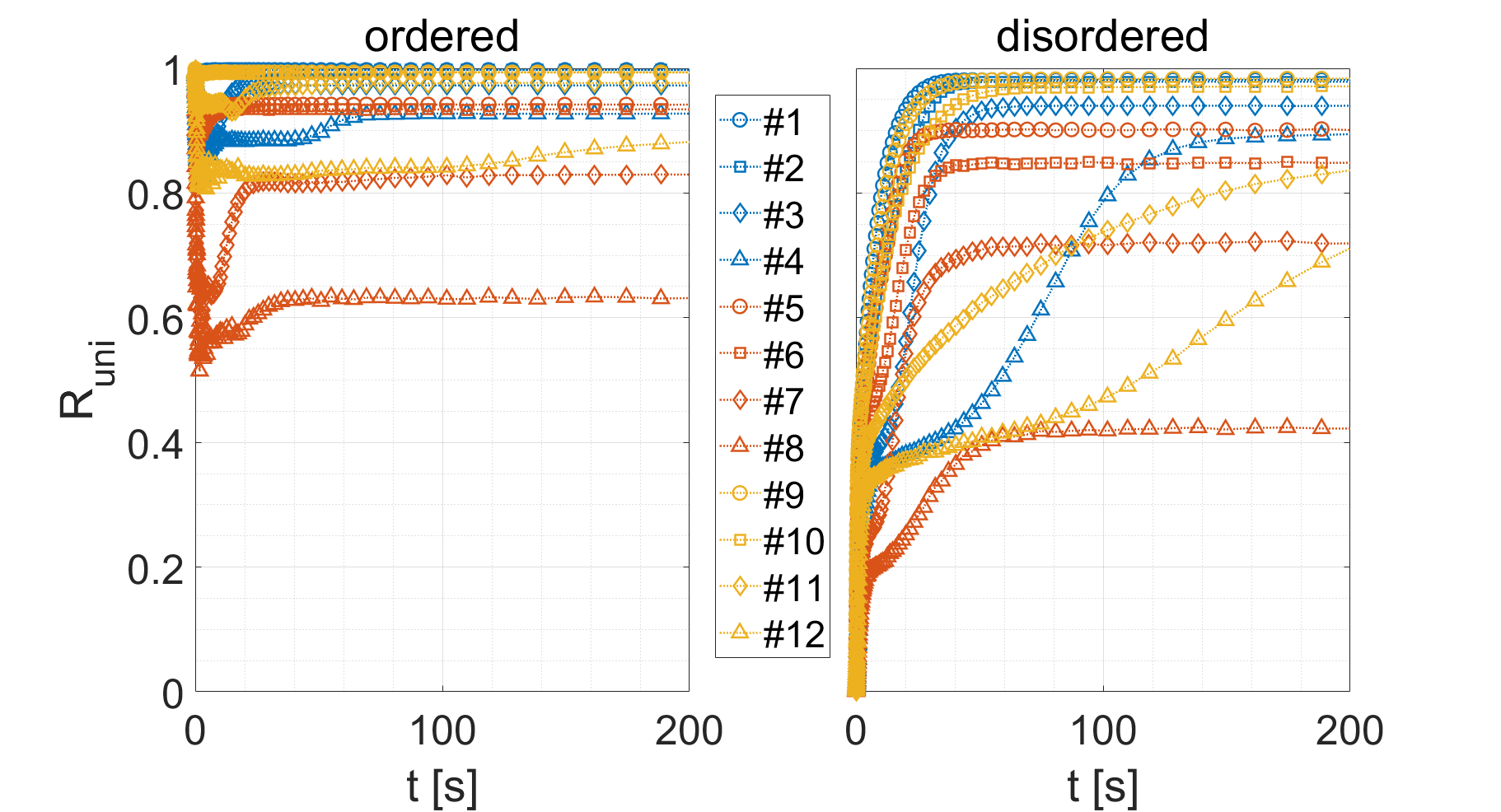}
    \caption{Evolution of $R_{uni}$ for phase ordered and disordered initial conditions at $\lambda=0.5$.}
    \label{fig:R_uni_therm}
\end{figure}

\end{document}